\documentclass[letterpaper]{article} 
\usepackage{aaai25}  
\usepackage{times}  
\usepackage{helvet}  
\usepackage{courier}  
\usepackage[hyphens]{url}  
\usepackage{graphicx} 
\usepackage{natbib}  
\usepackage{caption} 
\usepackage{algorithm}
\usepackage[noend]{algorithmic}

\usepackage{amsmath} 
\usepackage{amssymb}  
\usepackage{bm}
\usepackage{multirow}
\usepackage{tabularx}
\newtheorem{assumption}{Assumption}
\newtheorem{corollary}{Corollary}
\usepackage{booktabs}
\usepackage{subfig}

\usepackage{newfloat}
\usepackage{listings}
\DeclareCaptionStyle{ruled}{labelfont=normalfont,labelsep=colon,strut=off} 
\floatstyle{ruled}
\newfloat{listing}{tb}{lst}{}
\floatname{listing}{Listing}
\title{One for Dozens: \\Adaptive REcommendation for All Domains with Counterfactual Augmentation}
\author{
    Huishi Luo\textsuperscript{\rm 1}, 
    Yiwen Chen\textsuperscript{\rm 1}, 
    Yiqing Wu\textsuperscript{\rm 2}, 
    Fuzhen Zhuang\textsuperscript{\rm 1,3}\thanks{Corresponding author.}, 
    Deqing Wang\textsuperscript{\rm 3}
}
\affiliations{
    \textsuperscript{\rm 1} School of Artificial Intelligence, Beihang University\\
    \textsuperscript{\rm 2} Institute of Computing Technology, Chinese Academy of Sciences\\
    \textsuperscript{\rm 3} SKLSDE, School of Computer Science, Beihang University\\
    \{hsluo2000, yiwenchen\}@buaa.edu.cn, wuyiqing20s@ict.ac.cn, \{zhuangfuzhen, dqwang\}@buaa.edu.cn
}

\begin{document}

\maketitle

\begin{abstract}
Multi-domain recommendation (MDR) aims to enhance recommendation performance across various domains. However, real-world recommender systems in online platforms often need to handle dozens or even hundreds of domains, far exceeding the capabilities of traditional MDR algorithms, which typically focus on fewer than five domains. Key challenges include a substantial increase in parameter count, high maintenance costs, and intricate knowledge transfer patterns across domains. Furthermore, minor domains often suffer from data sparsity, leading to inadequate training in classical methods. To address these issues, we propose Adaptive REcommendation for All Domains with counterfactual augmentation (AREAD). AREAD employs a hierarchical structure with a limited number of expert networks at several layers, to effectively capture domain knowledge at different granularities. To adaptively capture the knowledge transfer pattern across domains, we generate and iteratively prune a hierarchical expert network selection mask for each domain during training. Additionally, counterfactual assumptions are used to augment data in minor domains, supporting their iterative mask pruning. Our experiments on two public datasets, each encompassing over twenty domains, demonstrate AREAD's effectiveness, especially in data-sparse domains. 
\end{abstract}

%
\begin{links}
    \link{Code}{https://github.com/Chrissie-Law/AREAD-Multi-Domain-Recommendation}
\end{links}

\section{Introduction}
\label{sec:intro}
Recommender systems (RSs) have become essential in many web applications to offer personalized recommendations and combat information overload. With an increasing variety of items, platforms like Amazon and Alibaba segment their offerings into different channels or content pages, necessitating that RSs efficiently manage recommendations across these diverse domains. Multi-domain recommendation (MDR) systems mitigate the high costs of maintaining separate models for each domain by capturing cross-domain user interests more effectively in a unified model, thus enhancing recommendation performance across various domains \cite{chang2023pepnet, li2023adl, zhao2023cross, zhang2023cctl, gan2024peace}.

\begin{figure}[t!]
    \centering
    \includegraphics[width=1.0\linewidth]{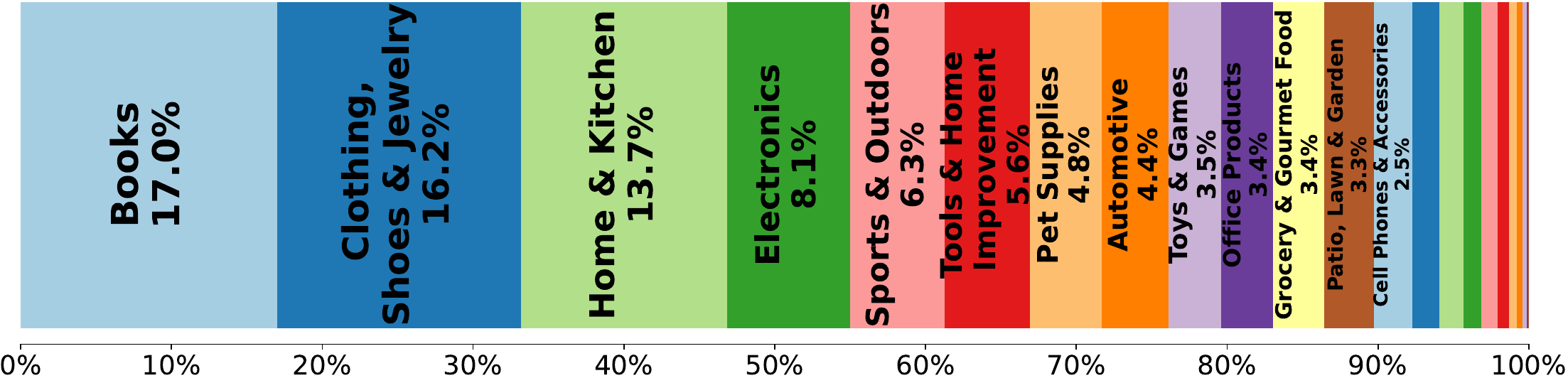}
    \caption{Sample size across 25 Amazon dataset domains, with 12 minor comprising less than 2\% of total samples.}
    \label{fig:amazon_domain}
\end{figure}

However, most MDR algorithms are limited to a handful of domains, typically fewer than five \cite{wang2022causalint, zhang2022sass, chang2023pepnet}. In contrast, real-world large-scale recommendation scenarios involve dozens or even hundreds of domains. For example, on Amazon\footnote{https://nijianmo.github.io/amazon/index.html}, there are as many as $25$ domains just identified using the coarsest item categorization (refer to Figure \ref{fig:amazon_domain}). Furthermore, these domains exhibit a significant long-tail effect, with major domains having substantially more data samples than minor ones. Therefore, training a multi-domain recommendation algorithm capable of handling such a diverse and extensive range of domain data faces the following challenges:

\textbf{Challenge 1: Scalability in Systems with Numerous Domains.} Traditional RS methods \cite{guo2017deepfm,song2019autoint,wang2021dcnv2} process multi-domain data in a single-domain manner, optimizing processing efficiency but sacrificing important domain specificity. To better capture the characteristics of each domain, most MDR algorithms refer to multi-task learning solutions, employing a separate “tower” network structure for each domain before output \cite{ma2018mmoe, sheng2021star, zou2022aesm, zhou2023hinet}. Nevertheless, this strategy significantly increases the model's parameter count and leads to inadequate convergence of these domain-specific networks as the number of domains grows. Additionally, pre-training-based MDR algorithms \cite{gu2021zeus, zhang2022sass} incur unacceptably high maintenance costs for fine-tuned models of each domain, and their direct application is often ineffective in domains with low cross-domain relevance.

\textbf{Challenge 2: Complex Cross-domain Knowledge Transfer Pattern.} Traditional MDR typically categorizes knowledge as either domain-shared or domain-specific. However, researchers \cite{standley2020tasks} highlight the directional nature of knowledge transfer, where beneficial knowledge in one domain may not be applicable in another, thus challenging this binary categorization framework. As domain numbers increase, the shortcomings of this framework become more evident, compounding the complexities of cross-domain knowledge transfer. Consequently, a crucial challenge in MDR with numerous domains is determining \textit{which domains should and should not be learned together}, a problem beyond the scope of traditional binary knowledge classification. While clustering domains and treating each cluster as a single domain offers a solution, this approach tends to overlook intra-cluster domain variations.

\textbf{Challenge 3: Large Variance in Sample Size across Domains.} As shown in Figure \ref{fig:amazon_domain}, nearly half of the domains account for less than 2\% of the total samples. This disparity leads to the dominance of data-abundant domains in the optimization process of existing MDR models \cite{zhang2023cctl}, impacting models regardless of their framework, whether those with explicit domain-specific parameters \cite{sheng2021star, zhou2023hinet} or those inspired by meta-learning \cite{zhu2021transfer, guan2022cross} and hypernetworks \cite{liu2023dtrn, chang2023pepnet}. Consequently, these models often suffer from insufficient optimization in sparse domains, which are particularly prevalent in datasets spanning a large number of domains.

To address the above challenges, we propose Adaptive REcommendation for All Domains with counterfactual augmentation (AREAD), a unified framework that adaptively models relationships across a vast array of domains at different granularities. AREAD is inspired by the hierarchical clustering technique and employs a hierarchical structure. Specifically, the structure utilizes a few expert networks at lower levels to capture coarse-grained domain knowledge, and a greater, yet still limited, number of expert networks at higher levels for finer-grained knowledge, thereby alleviating the parameter overhead in numerous domains (Challenge 1). AREAD leverages the Lottery Ticket Hypothesis \cite{frankle2019lottery}, generating and iteratively pruning a hierarchical expert selection mask for each domain during training, effectively capturing the complex knowledge transfer patterns across domains (Challenge 2). Furthermore, AREAD performs data augmentation for minor domains according to popularity-based counterfactual assumptions (Challenge 3). Finally, each domain utilizes its specific mask to determine the appropriate experts for the inference stage. We highlight our contributions as follows:

    $\bullet$ We propose the AREAD, addressing multi-domain recommendation involving dozens of domains.
    
    $\bullet$ AREAD incorporates a novel hierarchical expert sharing structure, with an iterative mask pruning approach to learn domain-specific sub-networks of experts for each domain. Additionally, we employ popularity-based counterfactual assumptions to augment data for minor domains.
    
    $\bullet$ Extensive experiments on two widely recognized public datasets, each with over 20 domains, demonstrate AREAD's significant improvements across diverse domains. AREAD excels in enhancing attention to minor domains while simultaneously improving overall performance.


\section{Methodology}
\label{sec:method}

\subsection{Preliminaries and Background}

Here we define the notations and problem settings of our study. Consider a set of recommendation domains $\mathcal{D}=\{1,2,\cdots,D\}$, with input features including common features $\bm{x}$ such as user historical behavior and context features, and a domain indicator $d \in \mathcal{D}$. The corresponding label $y(\bm{x},d) \in \{0,1\}$ represents whether a user has a positive interaction with an item in domain $d$.

In real-world recommendation platforms, the number of domains is often very large, and the sample size varies greatly among domains. We focus on cases where $D>20$. There exist domains $d_1, d_2 \in \mathcal{D}$ with a substantial disparity in their sample sizes, as expressed by the equation:
\begin{small}
\begin{equation}
    |\{(\bm{x},d,y)|d=d_1\}| \gg |\{(\bm{x},d,y)|d=d_2\}|.
\end{equation}
\end{small}
The optimization objective is formulated as follows:
\begin{small}
\begin{equation}
    \Theta=\underset{\Theta}{\arg \min } \sum_{d \in \mathcal{D}} \mathcal{L}(f(\bm{x},d), y(\bm{x},d)),    
\end{equation}
\end{small}
where $f$ is the MDR model and $\mathcal{L}$ denotes the loss function, 

\subsection{Overall Framework}
The overall structure of the AREAD framework encompasses three main parts: 

    $\bullet$ Hierarchical Expert Integration (HEI) is built upon the bottom Base Recommender, comprising multi-layer expert networks and gating units. It extracts and integrates domain knowledge of varying granularities to generate predictions, effectively reducing the parameter count by avoiding the need for separate output networks for each domain.
    
    $\bullet$ Hierarchical Expert Mask Pruning (HEMP) iteratively prunes the masks for selecting domain-specific experts, thereby identifying the most effective knowledge transfer patterns (Algorithm \ref{algo:hemp}). This process significantly reduces the time complexity of searching for the optimal transferable knowledge for the current domain from HEI.
    
    $\bullet$ Popularity-based Counterfactual Augmenter operates under the counterfactual assumption, suggesting that interactions with unpopular items in major domains are likely to occur similarly in minor domains. This is based on the rationale that if a user engages with a less popular item, the driving force is likely rooted in genuine preference rather than conformity, and such genuine preference remains unchanged across different domains.

\subsection{Hierarchical Expert Integration}

\subsubsection{Why HEI}
In MDR systems encompassing $D$ domains, existing models typically adopt a multi-task learning framework, constructing $D$ separate tower networks to compute domain-specific outputs. This architecture leads to a linear increase in the size of model parameters as $D$ grows. As for other strategies such as single-tower models, meta-learning, and hyper-network approaches, their training tends to be dominated by data from major domains, leading to insufficient optimization for minor domains. Moreover, these methods fail to explicitly capture the complex knowledge transfer pattern across multiple domains.

Considering the challenges posed by a large number of domains and the data sparsity issue in some of them, we believe that clustering domains is an effective strategy to reduce maintenance costs and ensure recommendation performance. However, \textit{how to measure the similarity between domains} and \textit{how to conduct clustering} are key problems to consider. For the former, the common practice is calculating similarity based on the output loss or gradient information of each domain \cite{bai2022saliency, wang2023exploration}. For the latter, using clustering algorithms to treat each cluster as a single domain for multi-domain learning is a feasible option. Nevertheless, these clustering methods exhibit three main drawbacks: 1) they are two-stage rather than end-to-end; 2) the calculation of similarity is inaccurate for minor domains due to insufficient optimization; and 3) intra-cluster domain individual characteristics are neglected.

\begin{figure}[t!]
    \centering
    \includegraphics[width=1.0\linewidth]{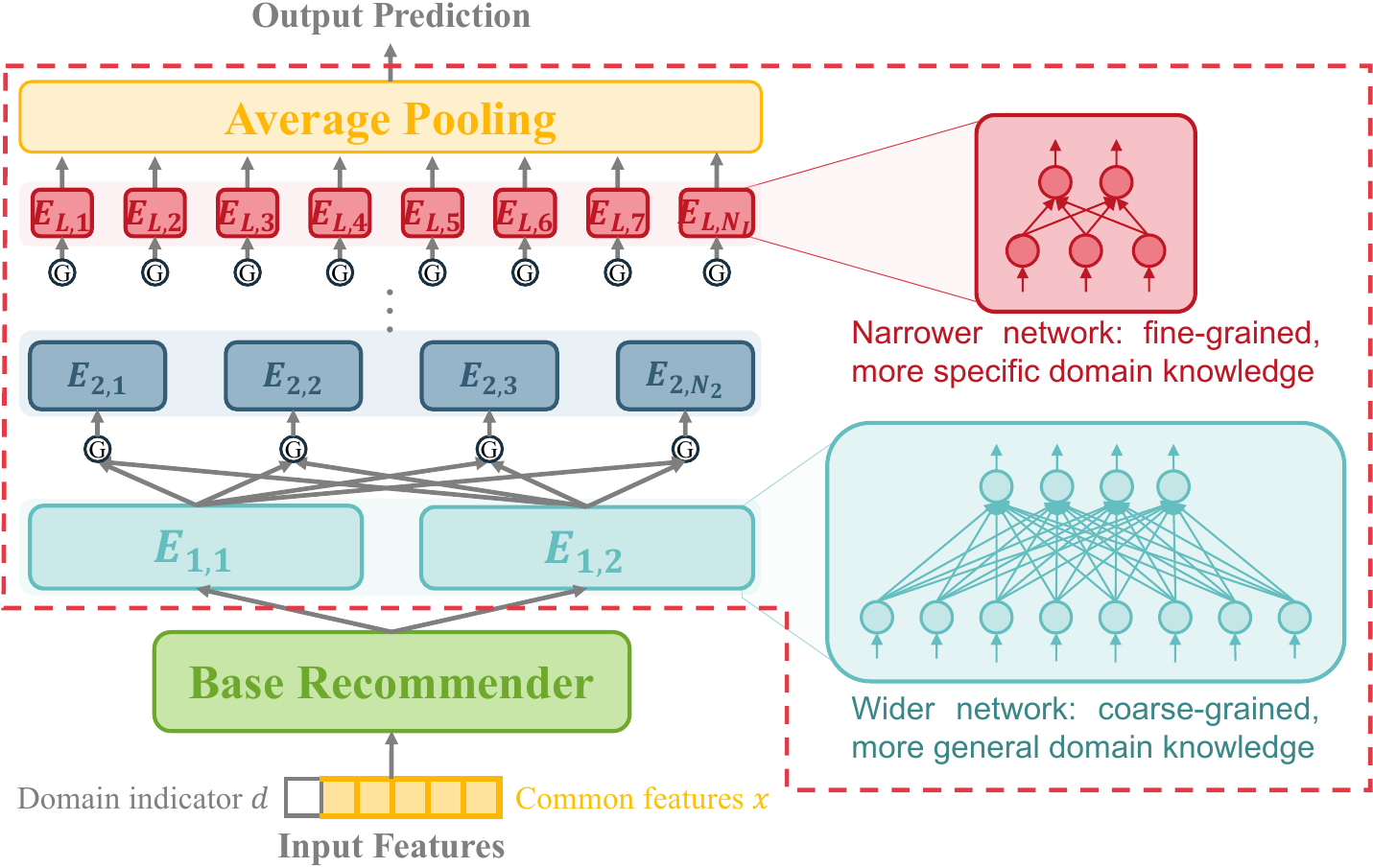}
    \caption{Hierarchical Expert Integration (HEI) uses multi-layer expert networks atop a base recommender to extract and integrate domain knowledge of varying granularities.}
    \label{fig:hei}
\end{figure}

\subsubsection{Main Description}
To tackle these clustering challenges, we propose a novel Hierarchical Expert Integration (HEI) module (Figure \ref{fig:hei}), inspired by hierarchical clustering. HEI is comprised of $L$ layers of expert networks and corresponding gating networks, in which each expert is implemented with a two-layer Multi-Layer Perceptron (MLP) and ReLU as the activation function. At the first level, HEI employs a small number of expert networks for coarse-grained clustering, where each expert captures knowledge of similar domains. To address more subtle differences within intra-cluster domains, we introduce higher layers of experts. As the number of layers increases, so does the number of experts per layer, yet still remaining significantly lower than the total number of domains $D$. These finer experts capture less knowledge, thus each network is structured to be narrower, requiring a reduced parameter count. Even with its multi-layered structure, HEI maintains a lower parameter count than assigning individual towers for each domain. 

The hierarchical structure of HEI ensures the extraction of knowledge across dozens of domains at varying granularities. The narrower experts in the higher layer support the adaptive, domain-specific learning of minor domains while minimizing interference from major domains. Moreover, the fully connected arrangement of expert networks between layers, combined with the subsequent Hierarchical Expert Mask Pruning, provides a more flexible exploration for learning each domain's transfer pattern.

\subsection{Hierarchical Expert Mask Pruning}
\label{sec:mask}

\subsubsection{Why HEMP}

\begin{figure}[t!]
    \centering
    \includegraphics[width=1.0\linewidth]{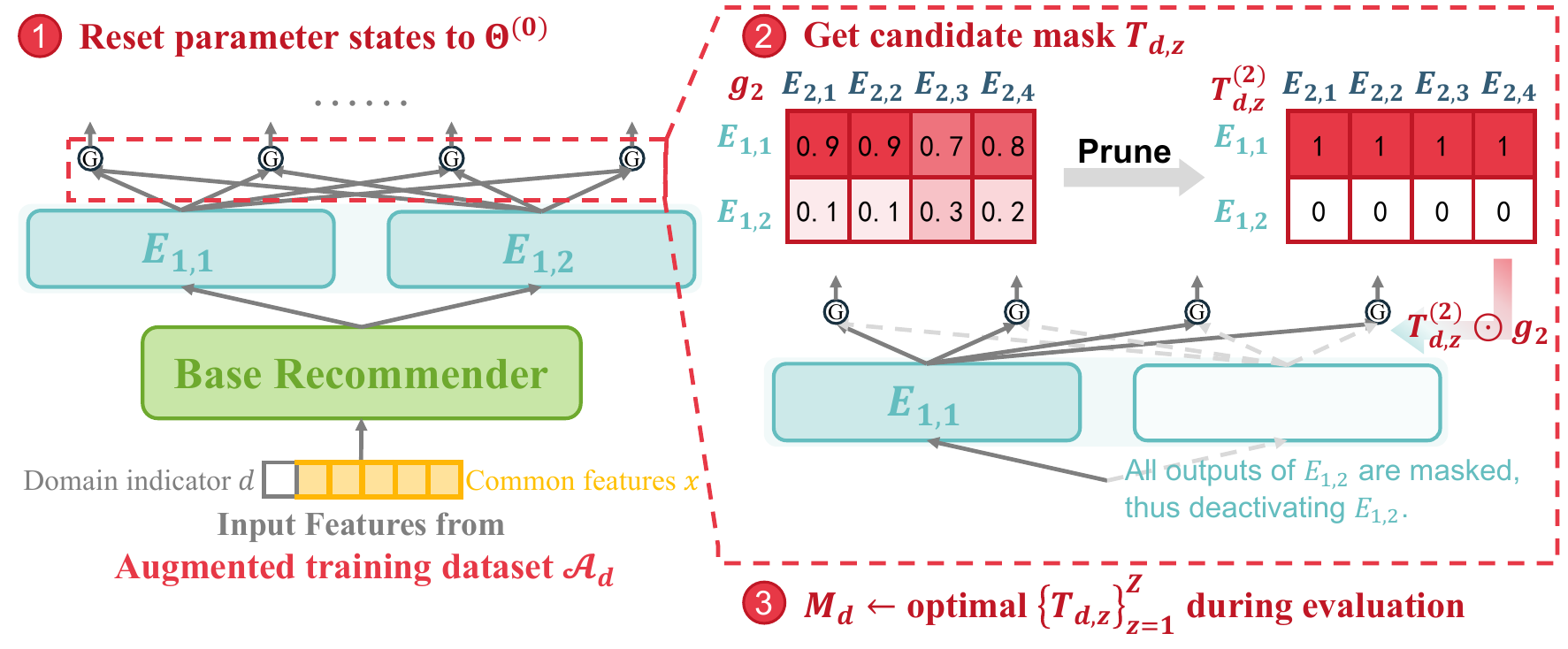}
    \caption{Hierarchical Expert Mask Pruning (HEMP) generates and iteratively prunes to select domain-specific experts.}
    \label{fig:hemp}
\end{figure}

For architecture like HEI with multiple experts, traditional multi-domain recommendation algorithms often explicitly dictate how parameters are shared, designating certain experts as domain-shared and others as domain-specific \cite{ma2018mmoe, tang2020ple, sheng2021star, shen2021sarnet, zhou2023hinet}. However, for a large number of domains, this binary division of domain knowledge into shared or specific falls short of capturing the complexities of knowledge transfer. Research by Standley et al. \cite{standley2020tasks} supports this, exploring the relationship between jointly trained tasks and searching for the best way to split five example tasks into groups for multi-task learning in experiments. This study reveals that the knowledge gained from jointly training certain task combinations can be more beneficial than completely shared knowledge across all tasks. Such insight is crucial for multi-domain learning involving dozens of domains, highlighting the importance of determining \textit{which domains should be jointly learned together and which should not}.

Directly searching for the optimal domains combination  for joint training is practically infeasible due to high complexity. Existing methods rely on gate networks, attention mechanisms, or hyper-networks to adaptively learn the importance of experts, an assumption that may be overly idealistic, especially with imbalanced data volumes across domains. AESM$^{2}$ \cite{zou2022aesm} provides an automatic expert selection framework, selecting the top $K$ most specific and most shared experts for the current domain at each layer based on their gate values. 
However, the greedy strategy in AESM$^{2}$, focusing on immediate layer-specific optimizations, may not always lead to globally optimal results because of the hierarchical knowledge transfer between layers.
\subsubsection{Main Description}

\begin{algorithm}[t]
\footnotesize
\caption{Hierarchical Expert Mask Pruning}
\label{algo:hemp}
\textbf{Input}: AREAD framework with parameter $\Theta$; pruning rate $\alpha$; initial sparsity threshold $S_0$; minimal sparsity $S$; augmented training datasets for $D$ domains $\{\mathcal{A}_1, \cdots, \mathcal{A}_D\}$; training datasets $\mathcal{T}$\\
\textbf{Output}: Optimized AREAD model; updated masks for each domain $\{M_d\}_{d\in\mathcal{D}}$
\begin{algorithmic}[1] 
\STATE Warm up AREAD with training data $\mathcal{T}$.
\WHILE{AREAD has not converged}
    \IF {condition to update domain masks is met}
        \STATE $\Theta^{(0)} \gets$ current parameter state of AREAD.
        \FOR{$d \in \mathcal{D}$}
            \FOR{$z \gets 1$ to $Z$}
                \STATE Reset $\Theta$ to $\Theta^{(0)}$
                \STATE Randomly initialize candidate mask $T_{d,z}$ based on gate values and initial sparsity threshold $S_0$.
                \FOR{$k$ steps}
                    \STATE Train AREAD with data sample from augmented training dataset. $\mathcal{A}_d$\ under the updating learning rate $lr_u$.
                    \STATE Prune $\alpha\%$ of gates with the lowest magnitudes from $[T_{d,z}^{(l)} \odot g_{l}, \text{ for } l = 2, \ldots, L]$.
                    \IF {$\frac{\|T_{d,z}\|_0}{\sum_{l=2}^{L}N_{l-1}N_l} \leq S$}
                        \STATE break.
                    \ENDIF
                \ENDFOR
                Update $M_d$ from $\{T_{d,z}\}_{z=1}^Z$
            \ENDFOR
        \ENDFOR
        \STATE Reset $\Theta$ to $\Theta^{(0)}$
    \ELSE
        \STATE Train AREAD with $\{M_d \text{ for } d \in \mathcal{D}\}$ and data sampled from $\mathcal{T}$\ under the learning rate $lr$.
    \ENDIF
\ENDWHILE
\end{algorithmic}
\end{algorithm}

Inspired by the Lottery Ticket Hypothesis \cite{frankle2019lottery}, we propose the Hierarchical Expert Mask Pruning (HEMP), a flexible architecture for selecting domain-specific experts (Figure \ref{fig:hemp}). The lottery hypothesis suggests that within a randomly initialized, dense neural network, there exists a subnetwork (referred to as “winning tickets”) that, if trained in isolation, can match the test accuracy of the original network in no more training iterations. Specifically, the winning tickets are identified via iterative pruning. Following this hypothesis, we posit that optimal networks of domain-specific experts exist within the HEI framework, and we utilize HEMP to generate hierarchical selection masks for experts specific to each domain. For domain $d$, its mask $M_d = [ M_d^{(l)} \in \{0,1\}^{N_{l-1} \times N_l} \, \text{for} \, l = 2, \cdots, L ]$ controls the use of the gating weight across hierarchical experts layers, with $N_l$ denoting the expert number in the $l$-th layer of HEI. Before the application of domain masks, the output of the $n$-th expert in the $l$-th layer, denoted $e_{l,n}$, is defined as:
\begin{small}
\begin{equation}
\label{eq:ori_e}
    e_{l,n} = \operatorname{MLP}_{\Theta_{l,n}}(\sum_{i=1}^{N_{l-1}} g_{l,n}[i]e_{l-1,i}),
\end{equation}
\end{small}
where $g_{l,n}\in[0,1]^{N_{l-1}}$ is the corresponding gating network output, and $g_{l,n}[i]$ is its $i$-th element, indicating the importance score for the $i$-th expert in the previous layer.
With domain masks applied, the output is reformulated as:
\begin{small}
\begin{equation}
\label{eq:e}
    e_{l,n} =\operatorname{MLP}_{\Theta_{l,n}}\left(\frac{\sum_{i=1}^{N_{l-1}} M_d^{(l)}[i,n]\,g_{l,n}[i]\,e_{l-1,i}}{\sum_{i=1}^{N_{l-1}} M_d^{(l)}[i,n] g_{l,n}}\right).
\end{equation}
\end{small}

In particular, with HEMP, the training of hierarchical experts encompasses two distinct phases: \textit{Warm-Up} and \textit{Training with Mask}. During the Warm-Up phase, the model undergoes preliminary training on the initialized hierarchical experts without the use of domain masks, to allow initial adaptation to data characteristics. In the subsequent Training with Mask phase, HEMP is applied independently to each domain to derive specific masks. This involves generating a set of $Z$ candidate temporary masks $T_{d,z}$ with initial sparsity $S_0$ for each domain and iteratively pruning these masks to achieve a certain sparsity threshold $S$ or a certain number of iterations. For candidate mask initialization, the vast space of potential random masks necessitates a balance between randomness and training efficiency. To achieve this, the initialization space is narrowed down based on previous gate values. Specifically, we calculate the average value of gate weights obtained during previous training steps on domain data, retain the top $S_0$ percent of largest gates, and further introduce randomness by inverting the masking state of some gates to generate an initial candidate mask for each domain. After these candidate masks are initialized, pruning occurs iteratively during training. $\alpha$ percent of the remaining gates with the lowest values are pruned from $[T_{d,z}^{(l)}\odot g_{l}\, \text{for} \, l = 2, \cdots, L ]$, where $g_{l}=[g_{l,1}, g_{l,2}, \cdots, g_{l,N_{l-1}}]$ and $\odot$ denotes element-wise multiplication. Upon acquiring $Z$ candidate masks $\{T_{d,z}\}_{z=1}^Z$, to select a single mask for subsequent training from these multiple candidates, we adhere to a straightforward principle: choosing the mask that demonstrates the best performance on domain data sampled from the training dataset $\mathcal{T}$. 
Note that all candidate masks for all domains start from a uniform parameter state for training and pruning to adhere to the lottery hypothesis. The process of searching for hierarchical expert selection masks is detailed in Algorithm \ref{algo:hemp}.

\subsection{Popularity-based Counterfactual Augmentation}

\begin{figure}[t!]
    \centering
    \includegraphics[width=1.0\linewidth]{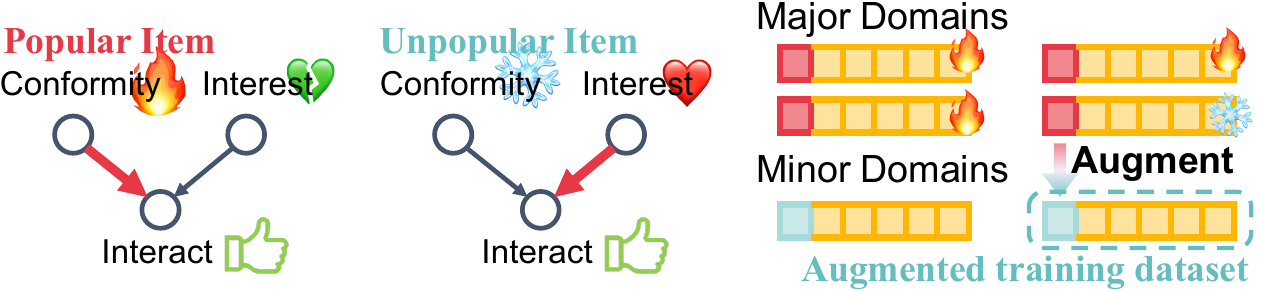}
    \caption{Popularity-based Counterfactual Augmenter utilizes counterfactual reasoning to infer genuine user interest across domains, augmenting interactions with unpopular items in major domains to minor domains.}
    \label{fig:aug}
\end{figure}

During the training of candidate masks in HEMP, minor domains struggle to support training for $k$ steps due to limited data, and repetitive sampling has little benefit in uncovering the characteristics of these domains. 
Inspired by Zheng et al.\cite{zheng2021disentangling}, we analyze the causal relationship between users' real interests and popularity, and on this basis, we perform counterfactual data augmentation\cite{ying2023camus, chen2024fairgap, chen2024fairdgcl}. For a popular item, a user might click on it simply because it has been clicked by many others, as seen on e-commerce platforms where items are often displayed with their sales figures. These interactions are primarily driven by user conformity rather than genuine interest. As a crucial factor for decision making, conformity describes how users tend to follow other people \cite{zheng2021disentangling}. Hence, we can break down the observed interactions into two user-side factors: \textit{interest} and \textit{conformity}. Their relationship can be depicted using a collider structure in a causal graph, expressed as $interest \rightarrow interact \leftarrow conformity$, meaning a positive interaction may result from either or both causes of interest and conformity. Based on this causal relationship, we propose the following assumption:
\begin{small}
\begin{assumption}
If a user has a positive interaction with a unpopular item, it is highly likely due to the genuine interest rather than the conformity.
\end{assumption}
\end{small}
This assumption, based on collider bias or the “explain-away” effect \cite{pearl2018book}, posits that a user does not need both conformity and interest to make a positive interaction; one is sufficient. Thus, with a positive outcome $y=1$, conformity and interest are spuriously negatively related. Consequently, when interacting with a unpopular item, the level of conformity is low, making it more likely that the interaction is driven by genuine preference, as illustrated in Figure \ref{fig:aug}. Additionally, we assume:
\begin{small}
\begin{assumption}
A user's genuine interests are consistent across domains.
\end{assumption}
\end{small}

Based on these, we derive a corresponding corollary to generate the augmented dataset $\{\mathcal{A}_1, \cdots, \mathcal{A}_D\}$:
\begin{small}
\begin{corollary}
If a user $u$ has a positive interaction $y=1$ with a unpopular item $i$ in a major domain, then it is also likely to occur in a minor domain:
\begin{small}
\begin{equation}
y(\bm{x},d\in\mathcal{D}_a|p(i)<\rho)=1 \implies y(\bm{x},d'\in\mathcal{D}_b|p(i)<\rho)=1,
\end{equation}
\end{small}
where $\mathcal{D}_a$ represents the set of major domains and $\mathcal{D}_b$ represents the set of minor domains. $p(i)$ means the popularity of $i$ correspond to $\bm{x}$ and $\rho$ is a certain popularity threshold.
\end{corollary}
\end{small}
For instance, a user purchasing niche philosophical books is likely to enjoy similarly niche philosophical art films. Although transferring items across domains can introduce noise, the augmentation's benefits significantly outweigh this noise at controlled augmentation ratios, as demonstrated in our hyper-parameter study. The strength of our counterfactual augmenter lies in its reliance on straightforward yet highly reasonable assumptions, enabling rapid augmentation implementation even in large-scale datasets.

\subsection{Model Optimization}
MMoE \cite{ma2018mmoe} serves as the base recommender system, facilitating shared feature interactions through embeddings corresponding to the input features. During the Warm-Up phase, data samples drawn from the training dataset are processed through both the foundational MMoE layer and the subsequent HEI layer, ultimately producing $N_L$ outputs. The average of these $N_L$ outputs is taken as the predicted value for the data sample: 
\begin{small}
\begin{equation}
\label{eq:warmup}
    \mathcal{L}(f(\bm{x},d),y(\bm{x},d)) = \mathcal{L}(\operatorname{Avg}\{\sigma(e_{L,i})\}_{i=1}^{N_L},y(\bm{x},d)),
\end{equation}
\end{small}
where $\sigma$ denotes the Sigmoid function. In the Training with Mask phase, for data $(x,d)$, the number of outputs calculated under the hierarchical expert selection masks $M_d$ is denoted by $|\mathcal{K}(M_d)|$, where:
\begin{small}
\begin{equation}
\mathcal{K}(M_d) = \{ j \in \{1, \cdots, N_L\} \mid \text{any}(M_d^{(L)}[:,j] > 0) \}
\end{equation}
\end{small}
To ensure that experts with smaller outputs receive adequate optimization, a strategy akin to Bagging \cite{kuncheva2014combining} in ensemble learning is adopted. During training, the loss between each output's prediction and the actual value is individually computed for gradient back-propagation:
\begin{small}
\begin{equation}
\label{eq:withmask}
    \mathcal{L}(f(\bm{x},d),y(\bm{x},d)) = \sum_{j\in \mathcal{K}(M_d)}\mathcal{L}(\sigma(e_{L,j}),y(\bm{x},d)).
\end{equation}
\end{small}
At inference, the outputs are averaged:
\begin{small}
\begin{equation}
    f(\bm{x},d) = \operatorname{Avg}\{\sigma(e_{L,j})\}_{j\in \mathcal{K}(M_d)}.
\end{equation}
\end{small}
The approach for utilizing candidate masks follows similarly, by substituting $M_d$ with $T_{d,z}$. In our experiments, the loss function $\mathcal{L}(\cdot)$ is set to binary cross-entropy loss.

\section{Experiments}
\label{sec:expe}

\begin{table}[t]
    \renewcommand\arraystretch{0.8}
    \footnotesize
    \centering
    \caption{Statistics of the datasets. The “Majority ratio” refers to the sample ratio in the largest domain, and “Minor domains” represent domains with less than 2\% of the samples.}
    \setlength{\tabcolsep}{6pt}{
    \begin{tabular}{crr}
    \toprule
    Datasets & Amazon & AliCCP \\
    \midrule
    \#Samples & 17,664,862  & 4,454,814  \\
    \#Positive samples & 8,790,575  & 329,416  \\
    \#Domains & 25    & 30 \\
    Majority ratio & 17\%  & 60.51\% \\
    \#Minor domains & 12    & 23 \\
    Augmentation ratio & 10\%  & 10\% \\
    \bottomrule
    \end{tabular}}%
    \label{tab:dataset}
\end{table}%

\subsection{Experimental Setup}

\begin{table*}[t]
    \renewcommand\arraystretch{0.8}
    \centering
    \caption{Performance comparison of different methods using five multi-domain metrics on Amazon and AliCCP datasets. Best and second-best results are highlighted in bold and underlined, respectively. $*$ indicates statistically significant differences (\textit{p}-value $< 0.01$) from the second-best result. Results are averaged over five runs.}
    \footnotesize
    \setlength{\tabcolsep}{1pt}{
    \begin{tabular}{ccccccccccc}
    \toprule
    \multirow{2}[4]{*}{Method} & \multicolumn{5}{c}{Amazon}            & \multicolumn{5}{c}{AliCCP} \\
\cmidrule(lr){2-6} \cmidrule(lr){7-11}          & AUC   & DomainAUC & Major5AUC & Minor10AUC & Minor5AUC & AUC   & DomainAUC & Major5AUC & Minor10AUC & Minor5AUC \\
    \midrule
    DeepFM & 0.7036  & 0.7111  & 0.7130  & 0.7251  & 0.7026  & 0.6003  & 0.5853  & 0.5869  & 0.5516  & 0.5669  \\
    DCN   & 0.7040  & 0.7114  & 0.7135  & 0.7256  & 0.6962  & 0.6084  & 0.5939  & 0.5952  & 0.5556  & 0.5709  \\
    AutoInt & 0.7039  & 0.7112  & 0.7132  & 0.7256  & 0.6954  & 0.6080  & 0.5930  & 0.5945  & 0.5520  & 0.5701  \\
    DCNv2 & 0.7040  & 0.7114  & 0.7134  & 0.7262  & 0.7009  & 0.6082  & 0.5939  & 0.5954  & 0.5540  & 0.5730  \\
    EPNet-S & 0.7033  & 0.7110  & 0.7131  & \underline{0.7267}  & 0.7026  & 0.6083  & 0.5946  & 0.5961  & 0.5531  & 0.5705  \\
    Isolated & -     & \textbf{0.7137 } & \textbf{0.7150 } & 0.7229  & \underline{0.7199}  & -     & 0.5851  & 0.5907  & 0.5536  & 0.5628  \\
    \midrule
    MMoE  & 0.7038  & 0.7113  & 0.7136  & 0.7255  & 0.7001  & \underline{0.6091}  & 0.5947  & 0.5963  & 0.5552  & 0.5733  \\
    PLE   & 0.7025  & 0.7099  & 0.7119  & 0.7248  & 0.6995  & 0.6090  & 0.5946  & 0.5959  & 0.5563  & 0.5765  \\
    STAR  & 0.7032  & 0.7032  & 0.7029  & 0.7042  & 0.6949  & 0.6075  & 0.6075  & 0.6078  & 0.6071  & 0.6036  \\
    AdaSparse & \underline{0.7053}  & 0.7102  & 0.7136  & 0.7258  & 0.6934  & 0.6074  & 0.5912  & 0.5927  & 0.5519  & 0.5695  \\
    HiNet & 0.7049  & 0.7109  & 0.7121  & 0.7247  & 0.6998  & 0.6075  & 0.5915  & 0.5928  & 0.5545  & 0.5738  \\
    EPNet & 0.7030  & 0.7106  & 0.7128  & 0.7251  & 0.6976  & 0.6084  & 0.5942  & 0.5957  & 0.5559  & 0.5769  \\
    PEPNet & 0.6995  & 0.7062  & 0.7073  & 0.7178  & 0.6891  & 0.6083  & 0.5941  & 0.5954  & 0.5584  & 0.5806  \\
    ADL   & 0.7033  & 0.7109  & 0.7130  & 0.7240  & 0.6991  & 0.6086  & \underline{0.6086}  & \underline{0.6087}  & \underline{0.6151}  & \underline{0.6131}  \\
    MAMDR & 0.6898 & 0.7079 & 0.7066 & 0.7194 & 0.6982 & 0.5957 & 0.5869 & 0.5869 & 0.5572 & 0.5691 \\
    \midrule
    AREAD & \textbf{0.7120*} & \underline{0.7131}  & \underline{0.7147}  & \textbf{0.7298* } & \textbf{0.7218* } & \textbf{0.6122* } & \textbf{0.6170* } & \textbf{0.6165* } & \textbf{0.6200* } & \textbf{0.6264* } \\
    \bottomrule
    \end{tabular}}%
    \label{tab:result}%
\end{table*}%

\subsubsection{Datasets}

We select two widely recognized datasets for our experiments: the \textbf{Amazon dataset}\cite{ni2019justifying} and the \textbf{AliCCP dataset}\cite{ma2018esmm}, both of which are organized by item category to define domains. Dataset statistics are detailed in Table \ref{tab:dataset}.

\subsubsection{Baselines}
\label{sec:baseline}
We compare AREAD with several state-of-the-art methods of both single-domain and multi-domain recommendation models. Since counterfactual data augmentation is not the main innovation of AREAD, we do not focus on comparing augmentation techniques. In practice, other suitable augmentation techniques can be used to enhance minor domain data. \textbf{(1) General Recommendation Models}: For single-domain recommendation contexts, we select well-regarded models such as \textbf{DeepFM} \cite{guo2017deepfm}, \textbf{DCN} \cite{wang2017dcn}, \textbf{AutoInt} \cite{song2019autoint} and \textbf{DCNv2} \cite{wang2021dcnv2}. \textbf{EPNet-S} first uses EPNet \cite{chang2023pepnet} to perform personalized selection on input embeddings, and shares one top DNN tower network across different domains. The \textbf{Isolated} model uses data solely from one domain for both training and testing, determining the optimal single-domain model and hyperparameters for each domain via grid search. \textbf{(2) Multi-Domain Recommendation Models}: These models, including \textbf{MMoE}  \cite{ma2018mmoe}, \textbf{PLE} \cite{tang2020ple}, \textbf{STAR} \cite{sheng2021star}, \textbf{AdaSparse} \cite{yang2022adasparse}, \textbf{HiNet} \cite{zhou2023hinet}, \textbf{EPNet} \cite{chang2023pepnet}, \textbf{PEPNet} \cite{chang2023pepnet}, \textbf{ADL} \cite{li2023adl}, and \textbf{MAMDR} \cite{luo2023mamdr}, are trained and tested on mixed multi-domain data. Due to the high cost of training separate-tower models across many domains, we pre-cluster domains and conduct multi-domain learning within each cluster as a single domain. ADL implements its own data clustering, and MAMDR, having validated domain scalability across numerous domains, both obviate the need for pre-clustering.

\subsubsection{Pre-Clustering Domains}
\label{sec:pre_clustering} 
We first train a single-domain DCN model using a combined dataset from all domains, chosen for its superior performance. Next, we record the distribution of losses for each domain on the test dataset and calculate the Kullback-Leibler (KL) divergence of these loss distributions to determine the distance between domains. Finally, we perform K-Means clustering \cite{macqueen1967kmeans} based on the distances computed among domains. For both datasets, the number of clusters is set to $3$. Note that the AREAD does not require pre-clustering.

\subsubsection{Metrics}
We utilize \textbf{AUC} \cite{fawcett2006introduction} for evaluation, a standard metric for binary classification tasks. In the context of multi-domain recommendation, where item interactions within each domain are independently assessed, aggregate AUC provides limited insights. Thus, we use \textbf{DomainAUC}, which measures the AUC for each domain separately, averaged and weighted by their sample sizes. Additionally, we assess performance with \textbf{Major5AUC}, \textbf{Minor10AUC}, and \textbf{Minor5AUC} to evaluate effectiveness in data-rich and data-sparse environments; these represent the weighted average AUCs of the largest and smallest domains, respectively. Notably, according to previous studies \cite{song2019autoint, jia2024d3}, an improvement in AUC at the \textbf{\underline{0.001} level (1\textperthousand)}  in CTR prediction tasks is considered significant and can lead to substantial commercial benefits online.

\begin{figure}[!t]
\centering
\includegraphics[width=\linewidth]{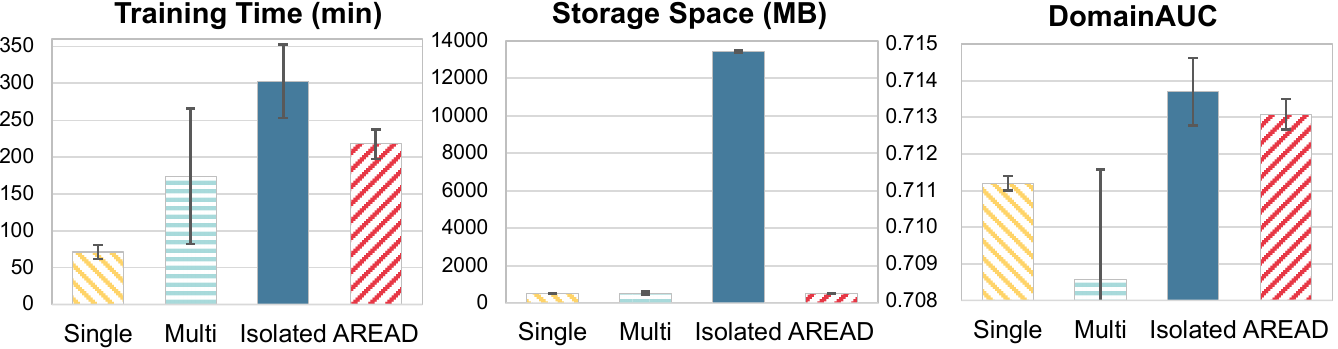}
\caption{Comparison of performance, storage space, and training time across Single-domain, Multi-domain models, Isolated method, and AREAD on the Amazon dataset.}
\label{fig:compare}
\end{figure}

\subsubsection{Implementation Details}
We use Adam \cite{kingma2014adam} optimizer with learning rates [$5e^{-4}$, $1e^{-3}$, $1.5e^{-3}$, $2e^{-3}$, $3e^{-3}$] and batch sizes [$1024$, $2048$, $4096$] optimized through grid search. 
Regularization coefficients are set at $1e^{-5}$. In AREAD, the Warm-Up phase trains $100$ batches. During the Train with Mask phase, masks update every $2000$ batches, exploring $Z=10$ candidate masks. Each mask $T_{d,z}$ is trained for $k=5$ batches on augmented data $\mathcal{A}_d$ of domain $d$. Initial mask sparsity is $S_0=0.7$, targeting $S=0.4$ with a pruning ratio of $\alpha=0.05$ per iteration. 

\subsection{Overall Performance}

Table \ref{tab:result} illustrates the experimental results of dozens of domains in the Amazon and AliCCP datasets. From the results, we have the following observations:

\begin{figure}[!t]
\centering
\includegraphics[width=1.0\linewidth]{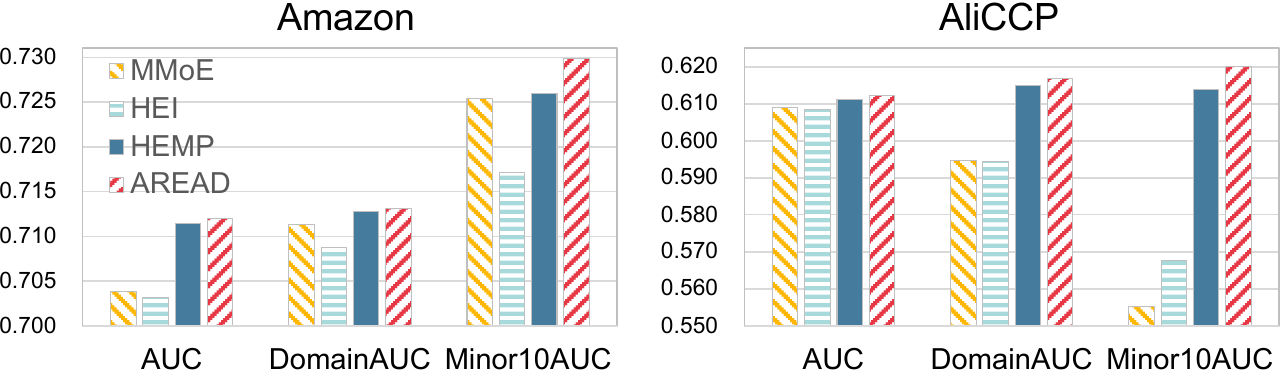}
\caption{Effectiveness of the sub-modules HEI, HEMP, and Counterfactual Augmentater in the AREAD framework.}
\label{fig:ablation}
\end{figure}

\begin{figure*}[!t]
    \centering
    \includegraphics[width=1.0\linewidth]{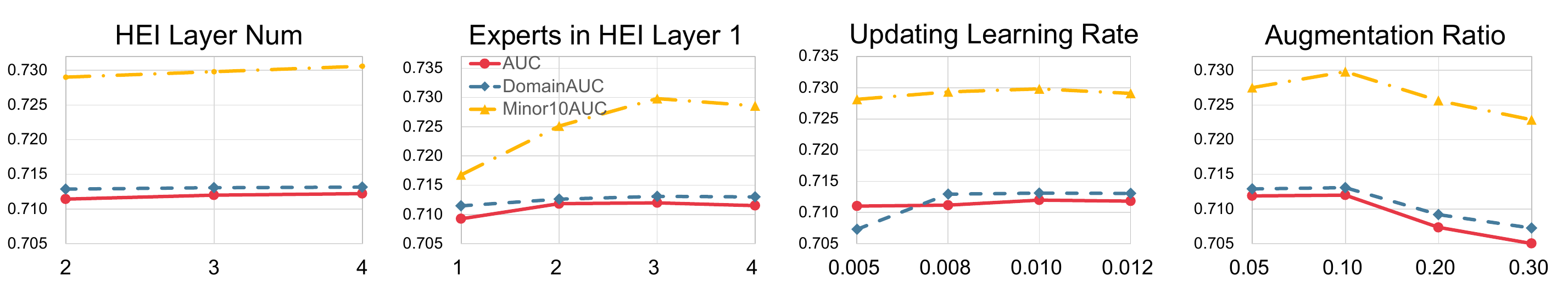}
    \caption{Hyperparameter study on Amazon dataset.}
    \label{fig:hyper}
\end{figure*}

(1) \textbf{The proposed method consistently achieves the best performance across general metrics.} AREAD significantly outperforms all baselines on nearly all metrics, with total AUC improvements of $6.7$\textperthousand{} on Amazon and $3$\textperthousand{} on AliCCP, both with \textit{p}-value $< 0.01$. More importantly, for the DomainAUC metric, which is of paramount interest in the multi-domain context, AREAD improves by $2$\textperthousand{} on Amazon and $8.4$\textperthousand{} on AliCCP compared to the best baseline trained on mixed-domain data. This underscores our framework's capability to effectively extract and leverage domain characteristics in recommendation scenarios with a large number of domains, leading to substantial enhancements in overall recommendation performance across all domains. 

(2) \textbf{Our model delivers outstanding results at acceptable maintenance costs.} On the Amazon dataset, AREAD excels in performance compared to all mixed-domain methods but does not surpass the “Isolated” method. However, the Isolated method incurs impractically high training and maintenance costs (see Figure \ref{fig:compare}), and its effectiveness decreases as domain similarities increase (see results on the AliCCP dataset in Table \ref{tab:result}). AREAD achieves an optimal balance between performance and cost, leading among existing models and closely variability the results of the Isolated method, even under significant domain variability.

(3) \textbf{Our model demonstrates exceptional performance on minor domains.} In the Amazon dataset, AREAD achieves significant improvements on Minor10AUC, with an increase of $3$\textperthousand{} (\textit{p}-value $< 0.01$) and $2$\textperthousand{}. In the AliCCP dataset, the corresponding improvements are $5$\textperthousand{} and $13$\textperthousand{}, respectively. These enhancements are attributable to the HEI and HEMP strategies within AREAD, which minimize interference from major domains and enhance the transfer of useful knowledge in HEI to minor domains. Additionally, counterfactual augmentation in AREAD addresses data scarcity in minor domains, further boosting performance.

\subsection{Ablation Study}
To further validate the effectiveness of the sub-modules in the AREAD framework, we evaluated the performance of the base recommender MMoE, the models with the HEI module, with the HEMP, and the complete model on both datasets, as shown in Figure \ref{fig:ablation}. Results indicate a slight performance decline when integrating the HEI module compared to the base recommender alone, confirming our previous hypothesis that without the use of expert selection masks, even a well-designed hierarchical expert architecture cannot ideally adapt and learn knowledge transfer patterns across dozens of domains. Integrating HEMP significantly enhances both AUC and DomainAUC, validating that domain masks substantially facilitate beneficial domain knowledge transfer. Further incorporation of the Popularity-based Counterfactual Augmentation results in notable improvements in the Minor10AUC metric, demonstrating that counterfactual data augmentation is particularly advantageous for the optimization of minor domains.

\subsection{Hyper-parameter Study}

To investigate the impact of various configurations within AREAD, we explored different hyperparameters: the number of layers ($L$) in the HEI module, the number of experts ($N_1$) in the first layer of HEI, the updating learning rate ($lr_u$) for model updates under candidate masks as specified in Algorithm \ref{algo:hemp}, and the percentage of augmented data ($r_{aug}$). Results on the Amazon dataset are illustrated in Figure \ref{fig:hyper}. 

We varied the number of layers $L$ in the HEI to $[2, 3, 4]$, with corresponding configurations for each layer's number of experts and hidden layer dimensions as follows: $[3(64,32), 6(16,8)]$, $[3(64,32), 6(32,16), 12(16,8)]$, and $[3(128,64), 6(64,32), 12(32,16), 24(16,8)]$. Experimental outcomes indicate that increasing the number of layers $L$ improves performance. Nonetheless, the incremental benefit of expanding $L$ from 3 to 4 layers is marginal. Consequently, we opted to maintain three layers in HEI, with each layer doubling the number of experts compared to the previous one. As $N_1$ ranged from $1$ to $4$, we observed that with fewer experts, AUC and DomainAUC performed well as the major domains dominated the optimization of HEI. However, performance in minor domains declined due to a lack of sufficient specific experts to integrate useful knowledge. 

The optimal $lr_u$ is $0.01$, with the metrics remaining relatively stable around this optimum. Lowering $lr_u$ diminishes the model’s capacity to quickly discern the effectiveness of candidate masks, adversely affecting DomainAUC. The augmentation ratio $r_{aug}$ directly influences the extent to which augmented data can affect model performance. Increasing $r_{aug}$ from $0.05$ to $0.1$ improves Minor10AUC without affecting AUC and DomainAUC. However, further increasing $r_{aug}$ introduces a substantial amount of noisy augmented samples, significantly impairing model performance.

\begin{figure}[!t]
    \centering
    \begin{minipage}[b]{0.49\linewidth} 
        \centering
        \includegraphics[height=0.11\textheight]{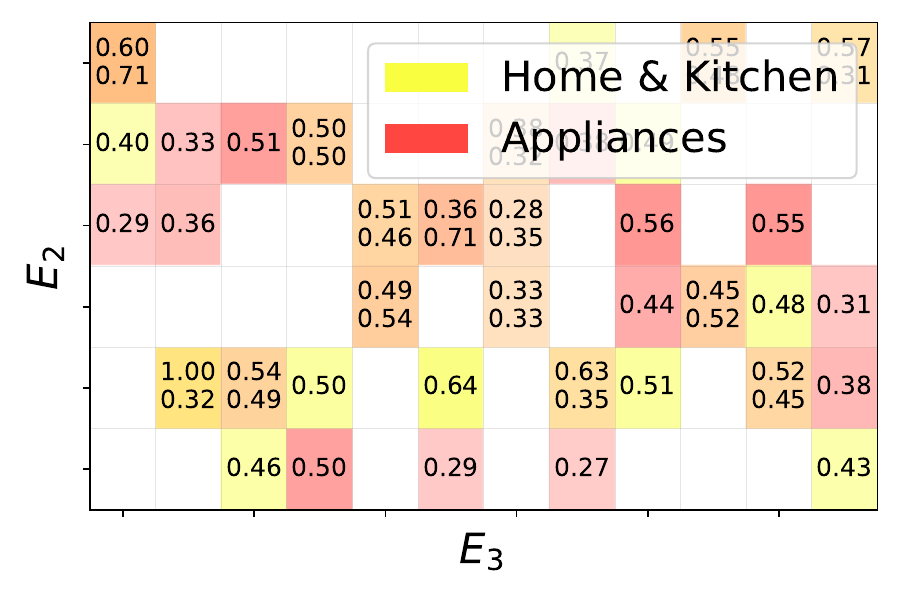}
    \end{minipage}
    \begin{minipage}[b]{0.49\linewidth}
        \centering
        \includegraphics[height=0.11\textheight]{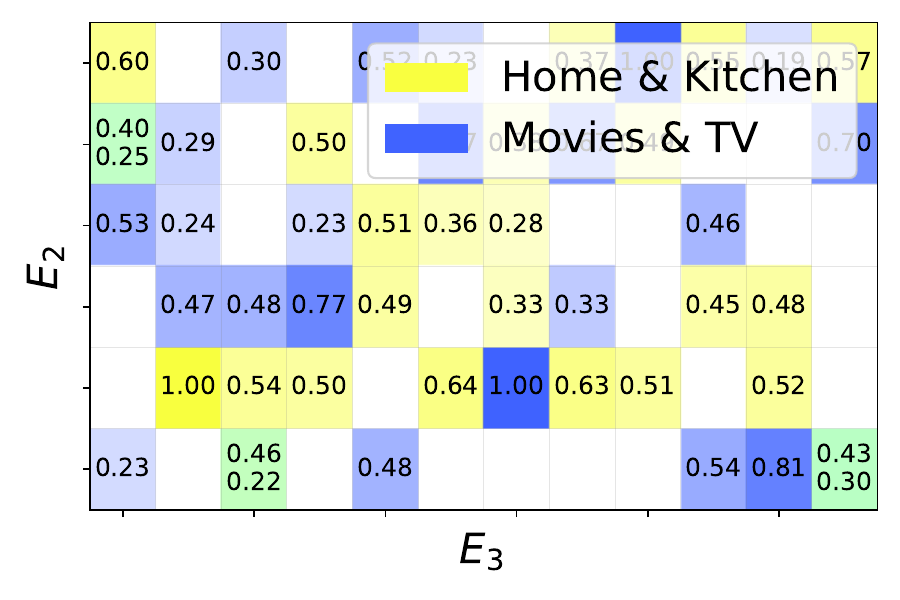}
    \end{minipage}
    \caption{Expert utilization in HEI of AREAD, with the left OR at 0.405 (“Home \& Kitchen” vs. “Appliances”), and the right OR at 0.065 (“Home \& Kitchen” vs. “Movies \& TV”).}
    \label{fig:case_mask}
\end{figure}

\subsection{Mask Analysis}

To explore AREAD's expert integration across domains, we analyze expert utilization on the Amazon dataset. AREAD uses three layers with $[3, 6, 12]$ experts each in HEI, with a target sparsity of $S=0.4$ in HEMP. After training on $25$ domains, we focus on “Home \& Kitchen” (HK), “Appliances” (AP), and “Movies \& TV” (MT), which constitute $0.137$, $0.018$, and $0.004$ of the dataset's total volume, respectively.

Figure \ref{fig:case_mask} shows the average gate weights between the second and third layers for the domain pairs “HK vs. AP” and “HK vs. MT”. We calculate the Mask Overlap Ratio (OR) to analyze domain relatedness \cite{sun2020learning}: $\mathrm{OR}\left(M_{d_1}^{(l)},M_{d_2}^{(l)}\right)={\left\|M_{d_1}^{(l)}\cap M_{d_2}^{(l)}\right\|_{0}}/{\left\|M_{d_1}^{(l)}\cup M_{d_2}^{(l)}\right\|_{0}}$. The OR for “HK vs. AP” is $0.405$, notably higher than the random overlap of $0.25$, highlighting similar functional features. In contrast, the OR for “HK vs. MT” is only $0.065$, indicating their relative non-relevance. Interestingly, despite the high similarity, HK and AP are often assigned to different groups in the pre-clustering stage. This underscores AREAD's ability to capture complex knowledge transfer patterns tailored to specific domain characteristics, which traditional methods might overlook. Furthermore, despite data volume disparities in the two smaller domains, the interpretability of expert utilization confirms both the reliability of Popularity-based Counterfactual Augmentation and AREAD's robust learning in minor domains.

\section{Related Work}
\label{sec:related_work}

Multi-domain recommendation systems aim to utilize diverse domain data to boost performance across all included domains. Inspired by multi-task learning \cite{ma2018mmoe,tang2020ple}, models often feature a domain-general base and domain-specific tower networks \cite{ma2018mmoe,tang2020ple,sheng2021star,shen2021sarnet,zou2022aesm,zhou2023hinet,chang2023pepnet}. This architecture is supplemented by gating \cite{ma2018mmoe,zou2022aesm,chang2023pepnet}, attention mechanisms \cite{shen2021sarnet}, hypernetworks \cite{chang2023pepnet, liu2023dtrn}, and dynamic weight parameters \cite{zhang2022m2m,yan2022apg}. Another approach, “Pre-Training + Fine-Tuning” \cite{gu2021zeus, zhang2022sass, luo2023mamdr}, enables cross-domain knowledge transfer through temporal parameter inheritance. Nonetheless, these models typically handle few domains and struggle to scale with the increasing number of real-world domains. A recent study \cite{li2023adl} has attempted to cluster scenes into groups for multi-domain adaptation, but this often results in suboptimal performance due to overlooked intra-cluster variations. Moreover, although MAMDR \cite{luo2023mamdr} has proven scalable across a vast number of domains in an industry-level system, its effectiveness may decline sharply in domains with significant differences, since it requires fine-tuning parameters for each domain from a unified shared model.

\section{Conclusion}
In this paper, we explore the issue of multi-domain recommendation across dozens of domains, and propose the Adaptive REcommendation for All Domains with counterfactual augmentation (AREAD) framework. AREAD employs Hierarchical Expert Integration to capture domain transfer knowledge at varying granularities within hierarchical expert networks. Building on this, the framework utilizes the Hierarchical Expert Mask Pruning algorithm to learn knowledge transfer patterns across numerous domains. Moreover, Popularity-based Counterfactual Augmentation is adopted to augment data for minor domains. Finally, experiments conducted on two public datasets, each encompassing over twenty domains, confirm the effectiveness of our approach.

\section*{Acknowledgements}
The research work is supported by the National Key Research and Development Program of China under Grant No. 2021ZD0113602, the National Natural Science Foundation of China under Grant Nos. 62176014 and 62276015, and the Fundamental Research Funds for the Central Universities.

\bibliography{aaai25}
\end{document}